\documentclass[11pt,twocolumn]{article}

\usepackage[a4paper,margin=1.8cm,columnsep=0.6cm]{geometry}
\usepackage{graphicx}
\usepackage{float}
\usepackage[most]{tcolorbox}

\usepackage{array}
\usepackage{booktabs}
\usepackage{caption}
\usepackage[space]{grffile}
\usepackage{latexsym}
\usepackage{textcomp}
\usepackage{longtable}
\usepackage{tabulary}
\usepackage{booktabs,array,multirow}
\usepackage{amsfonts,amsmath,amssymb}
\usepackage{natbib}
\usepackage{url}
\usepackage{hyperref}
\hypersetup{colorlinks=false,pdfborder={0 0 0}}
\usepackage{etoolbox}
\usepackage{acronym}
\usepackage{arydshln}
\newif\iflatexml\latexmlfalse

\AtBeginDocument{\DeclareGraphicsExtensions{.pdf,.PDF,.eps,.EPS,.png,.PNG,.tif,.TIF,.jpg,.JPG,.jpeg,.JPEG}}

\usepackage[utf8]{inputenc}
\usepackage[english]{babel}
\usepackage{multicol}

\usepackage[shortcuts]{extdash}

\usepackage{siunitx}
\usepackage{adjustbox}
\usepackage{xltabular}
\usepackage{subcaption}

\usepackage{authblk}

\graphicspath{{figures/}}

\newenvironment{widelongtable}
  {\onecolumn}
  {\twocolumn}

\title{Automated SysML-Based Verification of Discipline-Specific Models}

\author[1,2]{Daniel Marley}
\author[2]{Siyuan Ji\thanks{Corresponding author: s.ji@lboro.ac.uk}}
\affil[1]{MBDA Systems, Six Hills Way, Stevenage, UK, SG1 2DA}
\affil[2]{Wolfson School of Mechanical, Electrical and Manufacturing Engineering, Loughborough University, Epinal Way, Loughborough, UK, LE11 3TU}

\date{}

\begin{document}

\twocolumn[%
  \begin{@twocolumnfalse}
    \maketitle
    \begin{abstract}
    Current examples of SysML-based verification of discipline-specific models in the literature typically have two flaws. Firstly, they are developed in a tool-specific manner using proprietary APIs, limiting portability. Secondly, they focus on performance properties modelled via parametric diagrams, overlooking behavioural and interface properties that also require verification. This project addresses the problem with a verification process tailored to model-based verification, informed by common SysML tool capabilities and the UML Testing Profile, that enables automated verification of discipline\-/specific models from SysML test cases and returns the results to the SysML model for traceability. A mixed-method approach combining literature research and stakeholder interviews was used to derive validated stakeholder needs, which drove the specification and design of the process. The process was demonstrated end-to-end in two independent SysML tool-chains to evidence tool-agnosticism, and was shown to verify behavioural and interface requirements, including ordering, timing, and state-based responses, using SysML behavioural diagram constructs that parametric approaches alone cannot address.
    \end{abstract}
    \vspace{0.6em}
    \vspace{1.5em}
  \end{@twocolumnfalse}
]

\section{Introduction}
\label{sec:introduction}

In systems development, a substantial proportion of rework cost arises from defects that originate in early lifecycle phases but are not discovered until later stages. \citet{Feiler2010System} estimated the relative cost of resolving defects as a function of where they were introduced and where they were detected, finding that the requirements and design phases are the dominant source of rework expenditure. Detecting such defects earlier in the lifecycle, therefore, represents a significant cost reduction opportunity. \ac{MBSE} is a well-established route to achieving earlier defect detection \citep{Madni2019Economic}, and one active area of investigation is the use of the \ac{SysML} to perform \ac{MBV}, in which SysML test cases drive automated verification against discipline\-/specific models of the system under test.

The existing literature on SysML-based verification exhibits two recurring limitations that constrain its practical utility. The first is that published demonstrations are typically developed around a single SysML tool, relying on proprietary \ac{API}s or vendor-specific Extensible Markup Language implementations to couple the SysML model with the discipline\-/specific model. This tool specificity limits portability and impedes collaborative use across industry teams that work in different SysML toolchains. The second limitation is that the demonstrated approaches have focused predominantly on performance properties captured through parametric diagrams \citep{Morkevicius2015approach,Wang2019ModelBased,Castellani2022Virtual}. System requirements, however, routinely express expectations about behavioural response to stimuli, including temporal ordering, timing constraints, and state-based be\-hav\-iour, that parametric diagrams are not designed to represent. Taken together, these two limitations mean that existing \ac{MBV} practice is both less portable and narrower in verification scope than is desirable for industrial applications. SysMLv2 may offer a longer-term path to portability through its standardised \ac{API} and textual notation \citep{Bajaj2022Systems}, but tool support for SysMLv2 remained immature at the time of this work, making a practical contribution within current SysMLv1 toolchains both timely and necessary.

This paper addresses both limitations. The aim is to develop a tool-agnostic process for the automated verification of discipline-specific models using test cases defined in SysML, with verification results returned to the SysML model and traceable to the originating requirements. The process is synthesised from ISO\slash IEC\slash IEEE~15288 \citep{ISOIECIEEE152882023Systems}, ISO/IEC/IEEE~24641 \citep{ISOIECIEEE246412023Systems}, and the INCOSE Systems Engineering Handbook \citep{INCOSE2023INCOSE}, and is supported by a SysML profile informed by the \ac{UML} Testing Profile \citep{Omg2020UML}. Both the process and the profile are designed around capabilities that are common across current SysMLv1 tools, rather than around features specific to any one vendor.

The contributions of this work are threefold. First, a synthesised \ac{MBV} process is defined that integrates relevant activities from the systems engineering and model-based verification standards into a coherent workflow for the automated verification of discipline-specific models. Second, a minimal SysML profile for \ac{MBV} is defined, implemented, and evaluated, with design decisions justified against both the capabilities common to current SysMLv1 tools and the precedent established by \ac{UTP}; the profile is intentionally simpler than existing \ac{UTP} implementations to reduce required tool-specific automation. Third, and most significantly, the verification scope of SysML-based \ac{MBV} is extended beyond the parametric focus of the current literature by demonstrating that SysML behavioural diagram constructs can be used to define test cases that verify ordering, timing, and state-based requirements within a single workflow, a capability that parametric diagrams alone cannot provide. The process and profile are demonstrated end-to-end in two independent SysML toolchains, \ac{Magic SoS} and IBM Engineering Systems Design Rhapsody, against a Simulink-based system under test, and evaluated against a set of stakeholder needs elicited and validated through industry interviews.

The remainder of the paper is organised as follows. Section~\ref{sec:methodology} describes the research design and the methods used to elicit and validate the stakeholder needs that define the objectives of the work. Section~\ref{sec:literature-review} reviews the relevant standards, the \ac{UTP}, and the current state of SysML-based \ac{MBV}, and presents the synthesised \ac{MBV} process. Section~\ref{sec:systems-engineering-dev} presents the systems engineering development, including the validated stakeholder needs, the concept selection, the logical architecture, the SysML profile, and the end-to-end demonstration in both toolchains. Section~\ref{sec:discussion} discusses the results and their implications for industrial adoption. Section~\ref{sec:conclusions} summarises the conclusions and identifies directions for further work.
\section{Methodology}
\label{sec:methodology}

\subsection{Research design}
\label{sec:research-design}

The research output is a designed artefact, specifically a tool-agnostic \ac{MBV} process and a supporting SysML profile, rather than a theory of behaviour or a measurement of an existing system. The work is therefore framed as \ac{DSR} following the process model of \citet{Peffers2007Design} and the guidelines of \citet{Hevner2004Design}, an established framework for engineering research in which knowledge is generated through the construction and evaluation of a purposeful artefact, with utility and rigour assessed against an identified problem and a validated set of objectives.

The six phases of the Peffers et al.\ process model structure the paper as follows. Problem identification and motivation are addressed in the Introduction (Section~\ref{sec:introduction}) and elaborated through the Related Work (Section~\ref{sec:literature-review}), which establishes the gap in current \ac{MBV} practice. The objectives of the solution are defined as a validated set of stakeholder needs, elicited and prioritised through the activities described in Section~\ref{sec:data-collection} and presented in Section~\ref{sec:systems-engineering-dev}. Design and development of the artefact, comprising the \ac{MBV} process and the SysML profile, are also presented in Section~\ref{sec:systems-engineering-dev}. Demonstration is achieved through end-to-end implementation in two independent SysML toolchains against a Simulink-based discipline-specific model. Evaluation is performed by tracing demonstration outcomes back to the prioritised stakeholder needs, with findings discussed in Section~\ref{sec:discussion}. Communication is fulfilled by this paper.

Within this \ac{DSR} framework, the project adopts a pragmatic stance and a mixed methods approach \citep{Leavy2022Research}, in which qualitative methods are used to define and validate the objectives of the artefact, and the artefact is then developed and assessed against those objectives. ISO\slash IEC\slash IEEE~15288 \citep{ISOIECIEEE152882023Systems} and the INCOSE Systems Engineering Handbook \citep{INCOSE2023INCOSE} function as input to the design of the artefact rather than as the research method itself; their role in shaping the artefact is described in Section~\ref{sec:systems-engineering-dev}.

\subsection{Data collection}
\label{sec:data-collection}

A literature review covering verification processes, requirements for test case transformation, model-based verification approaches, and model exchange methods was conducted to establish the state of the art and to inform the design of the artefact. The results are presented in Section~\ref{sec:literature-review}.

Semi-structured stakeholder interviews were held to elicit and then validate the stakeholder needs and use cases that define the objectives of the artefact. Three practitioners participated, two systems engineering specialists and one discipline-specific modelling specialist, all external to and independent of the project. The first round elicited candidate stakeholder needs and use cases; the second round reviewed and refined the emerging artefact specification against those needs. Stakeholder needs were prioritised using the MoSCoW method \citep{Dick2017Requirements,ISOIECIEEE291482018Systems}, with the prioritisation recorded as an attribute of each need. The combination of independent external stakeholders and iterative two-round review supports confidence in the validity of the elicited needs. The panel of three participants is nonetheless a limitation on the breadth of needs captured, and a larger and more varied sample would be required before the set could be regarded as complete rather than indicative.

\section{Related Work}
\label{sec:literature-review}

\subsection{Verification processes}
\label{sec:verification-processes}





\begin{table*}[t] 
\footnotesize 
\centering
\setlength{\tabcolsep}{4.0pt} 
\renewcommand{\arraystretch}{0.98} 

\newcommand{\tablistitem}{\textbullet\hspace{0.5em}}

\caption{Synthesised Input-Process-Output}\label{tab:synthesised-ipo}

\begin{tabular}{|p{0.30\textwidth}|p{0.3\textwidth}|p{0.3\textwidth}|}
\hline
\rule{0pt}{2.5ex}\rule[-1.0ex]{0pt}{0pt}\textbf{Input} & \textbf{Process} & \textbf{Output} \\
\hline

\tablistitem Verification criteria \newline
\tablistitem System requirements \newline
\tablistitem System architecture description \newline
\tablistitem System design description \newline
\tablistitem System interface definition &

\tablistitem Prepare for verification \newline
\tablistitem Perform verification \newline
\tablistitem Manage results of verification &

\tablistitem Verification procedure \newline
\tablistitem Verified system requirements \newline
\tablistitem Verified system architecture and design \newline
\tablistitem Traceability mapping \newline
\tablistitem Verification report \newline
\tablistitem Verification records/ artefacts \\
\hline
\end{tabular}
\end{table*}

The INCOSE Systems Engineering Handbook \citep{INCOSE2023INCOSE} and ISO/IEC/IEEE~15288 \citep{ISOIECIEEE152882023Systems} describe verification through a common three-phase structure of preparing for verification, performing verification, and managing the results, and were reviewed together given their close alignment. For an \ac{MBV} process, several elements from these standards fall outside scope: high-level verification strategy and lifecycle concepts belong to the enclosing verification process; requirements on enabling systems and integration elements are managed at a higher level; and configuration management is handled separately. Within the remaining in-scope activities, the standards require that what is to be verified is identified and related to the system model, that success criteria are defined and captured, that verification actions are scheduled and executed, and that results are recorded and returned to the requirement with traceability maintained. ISO/IEC/IEEE~15288 further requires the export of results to external formats, since verification evidence must be submitted to approval authorities that may require it alongside other sources \citep{ISOIECIEEE152882023Systems,INCOSE2023INCOSE}.

ISO/IEC/IEEE~24641 \citep{ISOIECIEEE246412023Systems} is specific to \ac{MBSE} and treats verification and validation as a combined process; the two are distinct knowledge-building activities \citep{Kannan2025Theory}, but their tooling overlaps in an \ac{MBV} context, making it particularly relevant to an \ac{MBV} context. Its process description is less detailed than the INCOSE \ac{SEH} on execution sched\-ul\-ing and readiness checking, but adds useful emphasis on the need to integrate system models with discipline-specific models and to return results to the systems architecture. A notable gap in ISO 24641 is that it does not specify that the discipline-specific model and the systems model must share an explicit interface definition, which is a prerequisite for automated data exchange between the two model types.

Drawing together the relevant activities from all three sources, the synthesised \ac{MBV} process organises into three phases, shown in Table ~\ref{tab:synthesised-ipo}. In the prepare phase: identify and relate the system requirements to be verified by \ac{MBV}, and capture the verification method and success criteria for each. In the perform phase: order, schedule, and execute the verification actions, with a readiness check before execution. In the manage results phase: record verification outputs, determine results against success criteria, return results to the originating requirements, export to external formats as required, and maintain traceability throughout.

\subsection{Requirements to test case transformation}
\label{sec:system-requirements-test-cases}



\begin{table*}[t] 
\footnotesize 
\centering
\setlength{\tabcolsep}{4.0pt} 
\renewcommand{\arraystretch}{0.98} 

\caption{\ac{EARS} patterns \citep{Mavin2009Easy}}\label{tab:ears-patterns}

\begin{tabular}{|p{0.22\textwidth}|p{0.70\textwidth}|}
\hline
\rule{0pt}{2.5ex}\rule[-1.0ex]{0pt}{0pt}\textbf{Syntax name} & \textbf{Syntax pattern} \\
\hline
Generic & $<$optional preconditions$>$ $<$optional trigger$>$ the $<$system name$>$ shall $<$system response$>$ \\
\hline
Ubiquitous & The $<$system name$>$ shall $<$system response$>$ \\
\hline
Event-driven & WHEN $<$optional preconditions$>$ $<$trigger$>$ the $<$system name$>$ shall $<$system response$>$ \\
\hline
Unwanted behaviours & IF $<$optional preconditions$>$ $<$trigger$>$, THEN the $<$system name$>$ shall $<$system response$>$ \\
\hline
State-driven & WHILE $<$in a specific state$>$ the $<$system name$>$ shall $<$system response$>$ \\
\hline
Optional features & WHERE $<$feature is included$>$ the $<$system name$>$ shall $<$system response$>$ \\
\hline
Complex & Complex requirements use multiple conditional clauses. \\
\hline
\end{tabular}
\end{table*}

\begin{table*}[ht!] 
\footnotesize 
\centering
\setlength{\tabcolsep}{4.0pt} 
\renewcommand{\arraystretch}{0.98} 

\caption{\ac{EARS} entities mapped to SysMLv1}\label{tab:ears-mapping}

\begin{tabular}{|p{0.2\textwidth}|p{0.28\textwidth}|p{0.45\textwidth}|}
\hline
\rule{0pt}{2.5ex}\rule[-1.0ex]{0pt}{0pt}\textbf{\ac{EARS} Entity} & \textbf{SysMLv1 Entities} & \textbf{Notes} \\
\hline

$<$System name$>$ & Block\textsuperscript{BDD}, InterfaceBlock\textsuperscript{BDD}, Port\textsuperscript{BDD}, ProxyPort\textsuperscript{BDD}, FullPort\textsuperscript{BDD} & Work by \citet{Yue2011systematic} only proposes the class type.It does not separate out interface requirements as a distinct type, but considers this aspect covered by a port typed by an interface block. \\
\hline

$<$System response$>$ (behaviour) & Message\textsuperscript{SD}, SendSignalAction\textsuperscript{AD}, ControlFlow\textsuperscript{AD}, ObjectFlow\textsuperscript{AD}, TimeEvent\textsuperscript{AD}, Operation, Action\textsuperscript{AD}, CallBehaviourAction\textsuperscript{AD} & \citet{Yue2011systematic} suggests operations for doing verbs, actions considered as they are similar. \\
\hline

$<$System response$>$ (constraint) & DurationObservation\textsuperscript{SD}, DurationConstraint\textsuperscript{SD}, TimeConstraint\textsuperscript{SD}, TimeObservation\textsuperscript{SD}, ConstraintBlock\textsuperscript{Par} & \citet{Yue2011systematic} has used duration constraints and messages, only demonstrating their example with a sequence diagram. Similar \ac{SD} and \ac{AD} elements have been identified based on this. The constraint block comes from examples in \citet{Morkevicius2015approach} and \citet{Castellani2022Virtual}. \\
\hline

WHERE\ldots$<$Feature is included$>$ & N/A & This will be excluded as system variants can be managed by more effective means. \\
\hline

WHEN\ldots/\,IF\ldots, THEN\ldots$<$optional precondition$>$\ldots & Attributes (types not listed) & Both \citet{Yue2011systematic} and \citet{Dick2017Requirements} use attributes to capture properties. \\
\hline

WHEN\ldots/\,IF\ldots, THEN\ldots$<$trigger$>$ & AcceptEventAction\textsuperscript{AD}, Message\textsuperscript{SD}, LostMessage\textsuperscript{SD}, FoundMessage\textsuperscript{SD}, CombinedFragment\textsuperscript{SD[Alt,Opt]} & In both \citet{Yue2011systematic} and \citet{Dick2017Requirements} messages can be system inputs or outputs. The lost and found messages then fit well for the unwanted behaviour specification. Combined fragments can also be considered, with \citet{Yue2011systematic} having used alternative and parallel. \\
\hline

WHILE\ldots$<$in a specific state$>$ & StateInvariant\textsuperscript{SD}, State\textsuperscript{STM} & \citet{Yue2011systematic} uses the state invariant on a lifeline for requirement applicability, with strict conditions effectively functioning as a trigger. \\
\hline
\end{tabular}
\end{table*}

Establishing a formal requirements syntax creates a systematic basis for mapping requirement text to systems architecture elements. This concept is developed in \citet{Dick2017Requirements} through the idea of requirement boilerplates, which relate syntax constructs to modelling language constructs. A systematic literature review by \citet{Yue2011systematic} identified sixteen transformation approaches from requirements to UML models and found that restricted natural language was often used to facilitate automation, though the rationale for the restrictions was rarely documented. Rather than adopting one of these restricted formalisms, the more widely adopted \ac{EARS} \citep{Mavin2009Easy} was selected for this work. \ac{EARS} has well-understood strengths and weaknesses and is in established industrial use, which makes it a suitable basis for a tool-agnostic approach. The \ac{EARS} patterns are summarised in Table~\ref{tab:ears-patterns}. \ac{EARS} does not formally distinguish constraints from system responses in its pattern; separating an explicit $<$optional constraint$>$ field would improve precision in verification contexts.

The mapping of \ac{EARS} entities to SysMLv1 elements used in this work is shown in Table~\ref{tab:ears-mapping}, derived from the UML mappings of \citet{Yue2011systematic} translated to their SysML equivalents. Structural association elements are excluded as \ac{MBV} is concerned with responses to inputs rather than structural definitions. The table shows that sequence diagrams offer the broadest single-diagram coverage of \ac{EARS} constructs; however, activity diagrams provide a more port\-able basis for test case definition given the inconsistent support for sequence diagram execution across SysMLv1 tools, as discussed further in Section~\ref{sec:mbv-utp}.

\subsection{Model-based verification using SysML and the UML Testing Profile}
\label{sec:mbv-utp}

The \ac{UTP} is the first standardised model-based language for verification and validation \citep{Omg2020UML}, developed with SysMLv1 integration in mind. SysMLv1 already shares several elements with \ac{UTP}: SysMLv1 test cases carry the same definition, and the \texttt{VerdictKind} enumeration is derived from the \ac{UTP} verdict data type, differing only in the absence of the \texttt{none} literal \citep{SystemsMo2024Systems,Omg2020UML}. The additional profile elements required to extend SysMLv1 toward full \ac{UTP} alignment are therefore minor, though component and configuration items require further customisation to support automation. Despite being the most standards-grounded approach available, \ac{UTP} 2 has seen limited uptake: \citet{Bernardino2017Systematic} found only three \ac{UTP} citations from sixteen UML-based testing papers, and a 2024 systematic review of 149 papers by \citet{Cederbladh2024Early} makes no mention of it. The three conference papers that have applied \ac{UTP} consistently report benefits over textual approaches, including automation of previously manual test script production \citep{Baker2007Early}, easier development and maintenance through architecture reuse \citep{Stefanescu2010Using,Andaloussi2006Test}, and improved readability for practitioners familiar with UML \citep{Andaloussi2006Test}. Test Conductor, an IBM Rhapsody addon built on \ac{UTP} \citep{IBMEngine2025IBM}, provides an additional concept of note; the witness scenario, which records execution as a sequence diagram showing actual against expected events and values, is a particularly useful diagnostic mechanism that informed the design of the profile in this work.

Table~\ref{tab:sysml-mbv} summarises a representative selection of SysML- and UML-based \ac{MBV} approaches from the literature, including UML examples because SysML is derived from UML and the UML literature is more extensive \citep{Bernardino2017Systematic}. Papers that did not perform verification with an external discipline-specific model were excluded.

\begin{table*}[t]
\footnotesize 
\centering
\setlength{\tabcolsep}{3.0pt} 
\renewcommand{\arraystretch}{0.98} 

\caption{SysML/UML based \ac{MBV} approaches reviewed}\label{tab:sysml-mbv}

\begin{tabular}{|p{0.46\textwidth}|
                >{\centering\arraybackslash}p{0.06\textwidth}|
                >{\centering\arraybackslash}p{0.06\textwidth}|
                >{\centering\arraybackslash}p{0.06\textwidth}|
                >{\centering\arraybackslash}p{0.06\textwidth}|
                p{0.18\textwidth}|}
\hline
\rule{0pt}{2.5ex}\rule[-1.0ex]{0pt}{0pt}\textbf{Paper} & \multicolumn{4}{c|}{\textbf{SysML Diagram}} & \textbf{Discipline-Specific Model} \\
\cline{2-5}
 & \textbf{AD} & \textbf{SD} & \textbf{STM} & \textbf{Par} & \\
\hline

Development of an automated MBT toolchain from UML/SysML models \citep{Lasalle2011Development} &  &  & \checkmark &  & MATLAB \\
\hline
A Model-Based V\&V Test Strategy Based on Emerging System Modeling Techniques \citep{Wang2019ModelBased} & \checkmark & \checkmark & \checkmark & \checkmark & Not specified \\
\hline
Virtual Testing Workflows Based on the Function-Oriented System Architecture in SysML: A Case Study in Wind Turbine Systems \citep{Castellani2022Virtual} &  &  &  & \checkmark & MATLAB \\
\hline
Early UML Model Testing using TTCN-3 and the \ac{UTP} \citep{Baker2007Early} &  & \checkmark &  &  & Unspecified code \\
\hline
Using the \ac{UTP} for enterprise service choreographies \citep{Stefanescu2010Using} &  & \checkmark &  &  & SAP \\
\hline
A Test Specification Method for Software Interoperability Tests in Offshore Scenarios: A Case Study \citep{Andaloussi2006Test} &  & \checkmark &  &  & Unspecified code \\
\hline
Modelling and Analysis of Unmanned Aerial Vehicle System Leveraging Systems Modeling Language (SysML) \citep{Hossain2022Modeling} & \checkmark &  &  & \checkmark & MATLAB, Python \\
\hline
Enabling Model Testing of Cyber-Physical Systems \citep{Gonzalez2018Enabling} & \checkmark &  &  &  & MATLAB \\
\hline
Optimisation Workflows for Linking \ac{MBSE} and Multidisciplinary Analysis and Optimisation (MDAO) \citep{Habermehl2022Optimization} & \checkmark &  &  & \checkmark\textsuperscript{1} & MATLAB \\
\hline
A Model-based Approach for Verification of the Large Lenslet Array Magellan Spectrograph (LLAMAS) \citep{Stenzel2024Modelbased} & \checkmark &  &  & \checkmark\textsuperscript{1} & Python \\
\hline
\multicolumn{6}{l}{\scriptsize\textsuperscript{1}Parametric diagrams used in adjacent works rather than the cited paper directly.} \\
\hline
\end{tabular}
\end{table*}

\begin{table*}[t!] 
\footnotesize 
\centering
\setlength{\tabcolsep}{3.0pt} 
\renewcommand{\arraystretch}{0.98} 

\caption{SysMLv1 tool data exchange capabilities}\label{tab:tool-capabilities}

\begin{tabular}{|p{0.14\textwidth}|
                p{0.14\textwidth}|
                >{\centering\arraybackslash}p{0.04\textwidth}|
                >{\centering\arraybackslash}p{0.04\textwidth}|
                >{\centering\arraybackslash}p{0.04\textwidth}|
                >{\centering\arraybackslash}p{0.04\textwidth}|
                >{\centering\arraybackslash}p{0.04\textwidth}|
                >{\centering\arraybackslash}p{0.04\textwidth}|
                >{\centering\arraybackslash}p{0.04\textwidth}|
                >{\centering\arraybackslash}p{0.04\textwidth}|
                p{0.18\textwidth}|}
\hline
\rule{0pt}{10ex}\rule[-4ex]{0pt}{0pt}\textbf{Tool} & 
\textbf{XMI Export} & 
\rotatebox{90}{\textbf{Tables/CSV}} & 
\rotatebox{90}{\textbf{\ac{FMI}}} & 
\rotatebox{90}{\textbf{SysPhS}} & 
\rotatebox{90}{\textbf{MATLAB Integration}} & 
\rotatebox{90}{\textbf{Code Generation}} & 
\rotatebox{90}{\textbf{Model Execution}} & 
\rotatebox{90}{\textbf{API}} & 
\rotatebox{90}{\textbf{DDS}} & 
\textbf{Source(s)} \\
\hline

Rhapsody (9.2) & UML: 2.1, 2.1, 2.3, 2.4.1 & Yes & V1.0 & -- & Yes & Yes & Yes & Yes & Yes & Tool access \\
\hline
Cameo & UML: Clean, 2.5; MOF XMI; Eclipse UML 2 XMI & Yes & V1.0, V2.0 & V1.0 & Yes & Yes & Yes & Yes & Yes & Tool access \\
\hline
MagicDraw / \ac{Magic SoS} & UML 2.5; MOF XMI; Eclipse UML 2 XMI & Yes & V1.0, V2.0 & -- & Yes & Yes & Yes\textsuperscript{2} & Yes & Yes\textsuperscript{3} & Tool access, \citep{Nigischer2021Multidomain} \\
\hline
Eclipse/Papyrus & XMI: 2.5.1 & Yes & V2.0 & -- & No & Yes & Yes & Yes & No & Tool access, \citep{Nigischer2021Multidomain} \\
\hline
Enterprise Architect (17.1) & UML: 2.1, 2.4, 2.4.2, 2.5.1 & Yes & -- & V1.0 & Yes & Yes & Yes & Yes & Yes &  \citep{Nigischer2021Multidomain,EnterpriseArchitectUserGuide} \\
\hline
PTC Modeller (10.1) & XMI 2.4 & Yes & -- & -- & Yes & Yes & Yes & Yes & -- & \citep{Nigischer2021Multidomain,OverviewoOverview} \\
\hline
Modelio (5.4.1) & EMF 3.0.0; UML: 2.1.1, 2.2, 2.3, 2.4.1 & No & -- & -- & No & Yes & No & Yes & No & Tool access, \citep{Nigischer2021Multidomain} \\
\hline
\multicolumn{11}{l}{\scriptsize\textsuperscript{2}Via Cameo Simulation Toolkit. \textsuperscript{3}Via DDS plugin.} \\
\hline
\end{tabular}
\end{table*}

Parametric diagrams dominate the SysML based approaches in Table~\ref{tab:sysml-mbv}, with activity diagrams appearing in a subset of examples and state machine diagrams used rarely. Sequence diagrams occur only once among the SysML examples because most SysML tools cannot drive simulation from sequence diagrams natively; \ac{UTP}-based approaches are more likely to include sequence diagram-based test definitions because their tooling is designed for software testing. A structural diagram is common across all approaches to define interfaces and test architecture.
Six of the ten papers used MagicDraw, which accounts for the concentration of MATLAB and Python integration methods that exploit MagicDraw's native connectors; where model transformation is used, the source is typically XMI. This tool concentration is a direct expression of the portability limitation identified in Section~\ref{sec:introduction}.


Two further observations from the literature inform specific design decisions in this work. \citet{Gonzalez2018Enabling} found that the standard SysML profile was insufficient for co-simulation with MATLAB and required additional stereotypes beyond \ac{UTP} coverage, illustrating the trade-off between a richer domain-specific profile and tool\-/specific requirements; this reinforces the case for a deliberately minimal profile. \citet{Lasalle2011Development} demonstrated that a restricted subset of UML \ac{STM}s, enforced by model checks, could form a usable testing language without requiring full language coverage, a principle that applies equally to the activity diagram approach used in this work.

\subsection{SysML tool capabilities}
\label{sec:sysml-tool-capabilities}

The portability of an \ac{MBV} approach depends not only on the logical design of the process and profile but also on which integration and execution capabilities are consistently available across SysMLv1 tools. Table~\ref{tab:tool-capabilities} compares seven representative tools. \ac{FMI} support, model execution, and \ac{API} access are the capabilities of primary interest, as they are the mechanisms by which a SysML model can drive and interrogate a discipline-specific model at runtime. The majority provide model execution and \ac{API} access. \ac{FMI} is supported by Cameo, MagicDraw/\ac{Magic SoS}, and Rhapsody, though Rhapsody supports only \ac{FMI} v1.0. XMI export is universal but, as noted above, vendor implementations vary in ways that limit its reliability as a portable exchange format.

\ac{FMI} emerges from Table~\ref{tab:tool-capabilities} as the most suitable model exchange mechanism for a tool agnostic approach, being an open standard with broad support across the surveyed tools and without dependence on any vendor-specific integration. The universal availability of \ac{API} access suggests that API-based result capture is a viable fallback for tools where native simulation result recording is unavailable. Both observations inform the concept selection in Section~\ref{sec:concept-selection}.

\subsection{Summary and gap}
\label{sec:literature-review-summary}

The review establishes that a coherent \ac{MBV} process can be synthesised from the relevant standards, that \ac{UTP} provides a principled basis for an \ac{MBV} domain-specific language, and that \ac{FMI} offers the most portable model exchange mechanism available across current SysMLv1 tools. The existing \ac{MBV} literature, however, is concentrated around \ac{Par} approaches and around MagicDraw\-/specific implementations. Behavioural diagram approaches exist but are rare in the SysML context and do not exploit the decision logic, time events, and timeout constructs needed to verify the full range of requirement types that \ac{EARS} can express. No published work has demonstrated a tool agnostic \ac{MBV} workflow that addresses both parametric and behavioural requirements within a single process. This gap is what the present work addresses.
\section{Systems engineering development}
\label{sec:systems-engineering-dev}

\subsection{Stakeholder needs and requirements definition}
\label{sec:stakeholder-needs-and-requirements-definition}

The stakeholder needs were derived from the literature review and refined through industry stakeholder interviews, as described in Section~\ref{sec:data-collection}. The validated and prioritised needs are shown in Table~\ref{tab:consolidated}, organised into six functional groups covering model customisation, test management, interface management, test case definition, verification execution, and recording of verification outputs. MoSCoW prioritisation was applied by the stakeholder interviewees, with the majority of the must-have needs clustered in test management, interface management, verification execution, and result recording, reflecting the stakeholders' primary concern with end-to-end traceability from requirement to verified result.

\begin{figure*}[t]  
  \centering
  \includegraphics[width=0.7\textwidth]{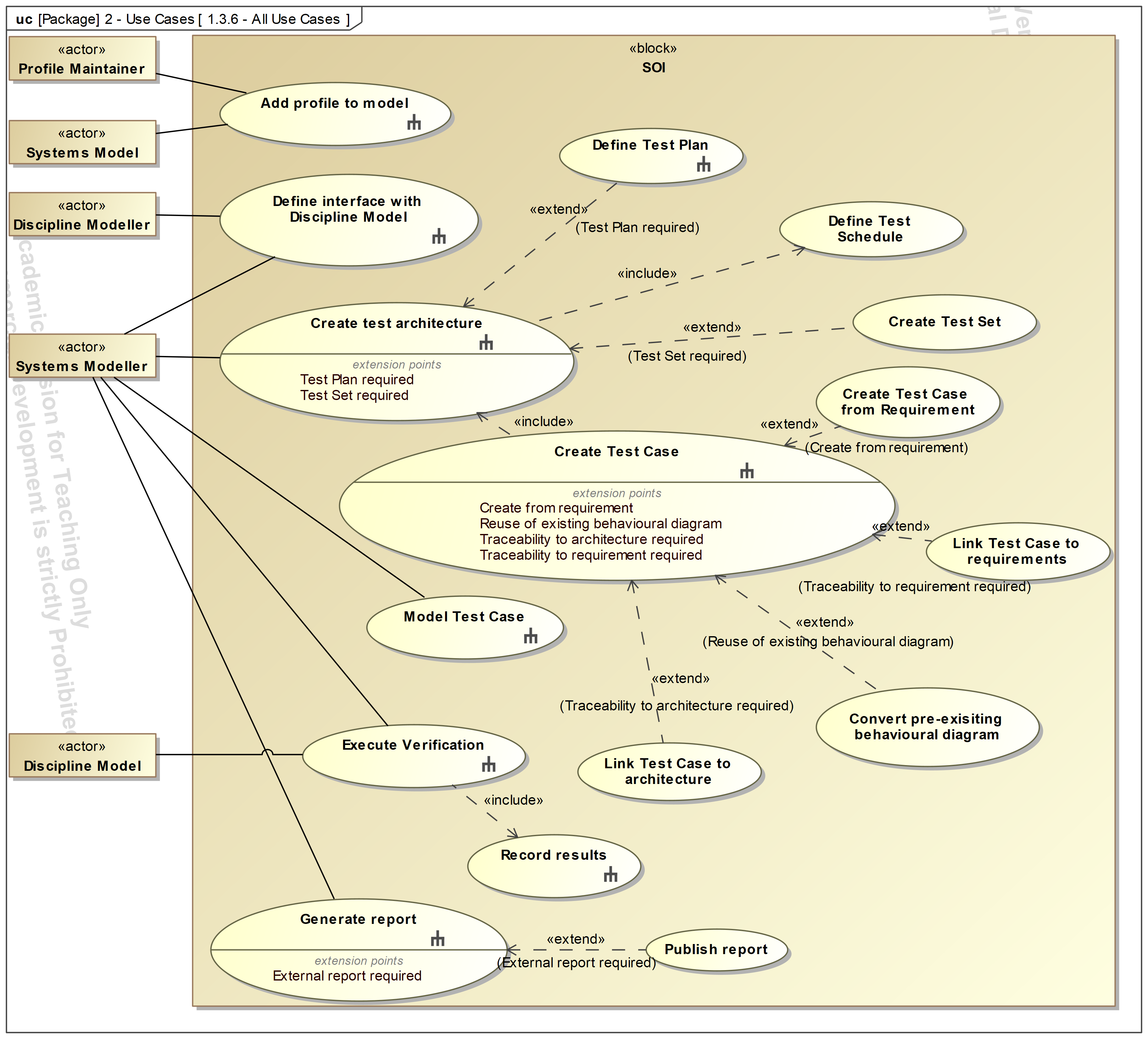} 
  \caption{\ac{UC} model showing the system of interest boundary and the actor blocks that interact with the \ac{MBV} capability}
  \label{fig:use-cases}
\end{figure*}

\ac{UC}s were captured to refine and bound the stakeholder needs, as shown in Figure~\ref{fig:use-cases}. The system of interest (SOI) boundary block represents the system under development, and the actors are modelled as blocks rather than conventional actor symbols to permit further specification in the architecture as required.

\subsection{Concept selection}
\label{sec:concept-selection}

A structured evaluation of candidate integration approaches was conducted, considering achievability, maintainability, tool coverage, and applicability across SysML versions, weight\-ed by stakeholder need prioritisation. SysMLv1 model execution combined with model exchange via the \ac{FMI} achieved the highest weighted score, as the tool capability survey in Section~\ref{sec:sysml-tool-capabilities} established that model execution and \ac{FMI} co-simulation are available across the majority of SysMLv1 tools. Any co-simulation exchange mechanism compatible with the tool is sufficient; API-based access to simulation values provides a fallback where a native exchange method is unavailable. SysMLv1 was preferred over SysMLv2 on the grounds of toolchain maturity at the time of this work; the concept is, however, considered transferable to SysMLv2, as discussed in Section~\ref{sec:discussion}.

\subsection{System architecture definition}
\label{sec:system-architecture-definition}

\begin{figure}[t]
  \centering
  \includegraphics[width=1\columnwidth]{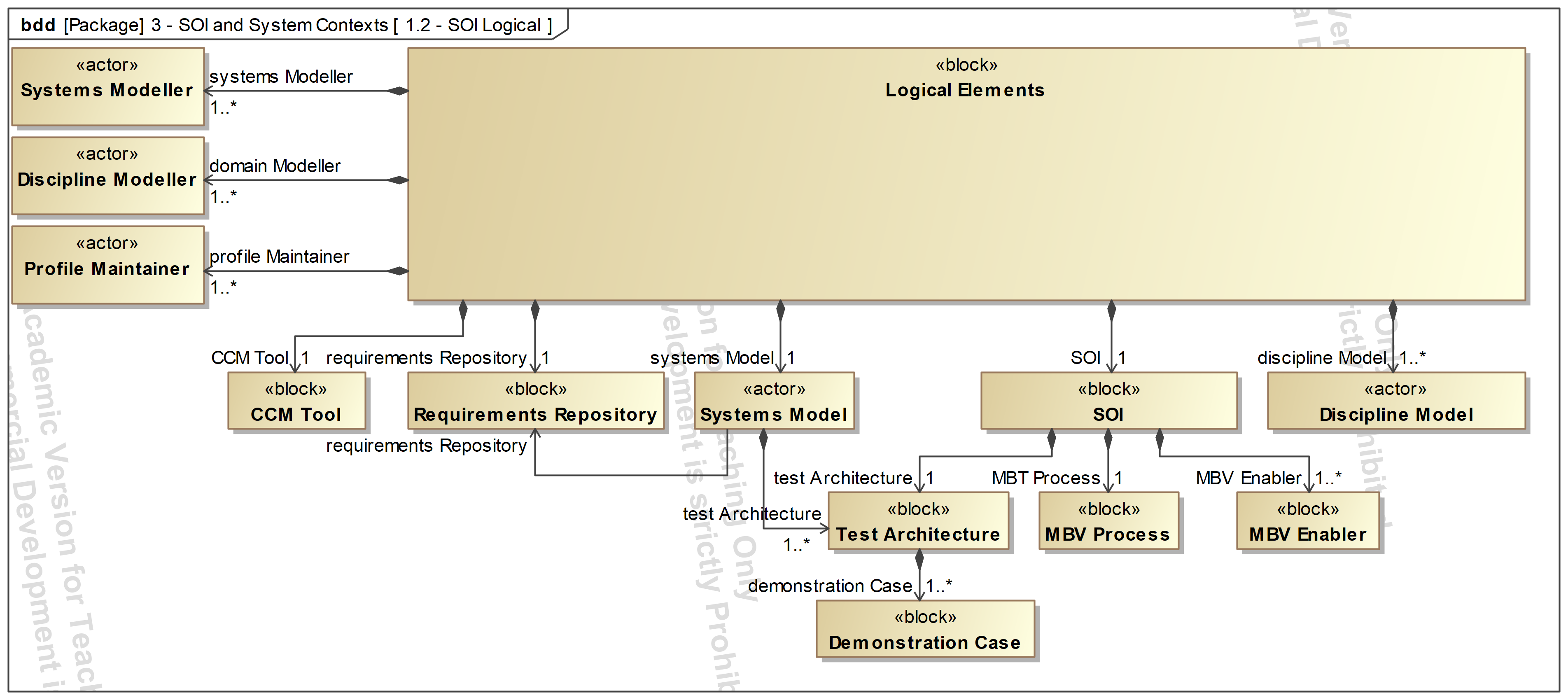}
  \caption{Context architecture \ac{BDD} showing the system of interest boundary, human actor roles, and peer context systems used to model the \ac{MBV} process}
  \label{fig:context-arch}
\end{figure}

The context architecture is represented as a \ac{BDD}, shown in Figure~\ref{fig:context-arch}. Actors interacting with the system appear on the left of the diagram and context systems at the bottom, giving a solution-agnostic view of the system boundary. The SOI block represents the system under development; peer context systems are modelled with blocks so that the \ac{MBV} process can be modelled in \ac{AD}s. Configuration and change management tools and requirements repositories were excluded from further modelling because their interactions with systems modelling tools are well understood and would not add value to the stakeholders of this work.

\begin{figure}[t]
  \centering
  \includegraphics[width=1\columnwidth]{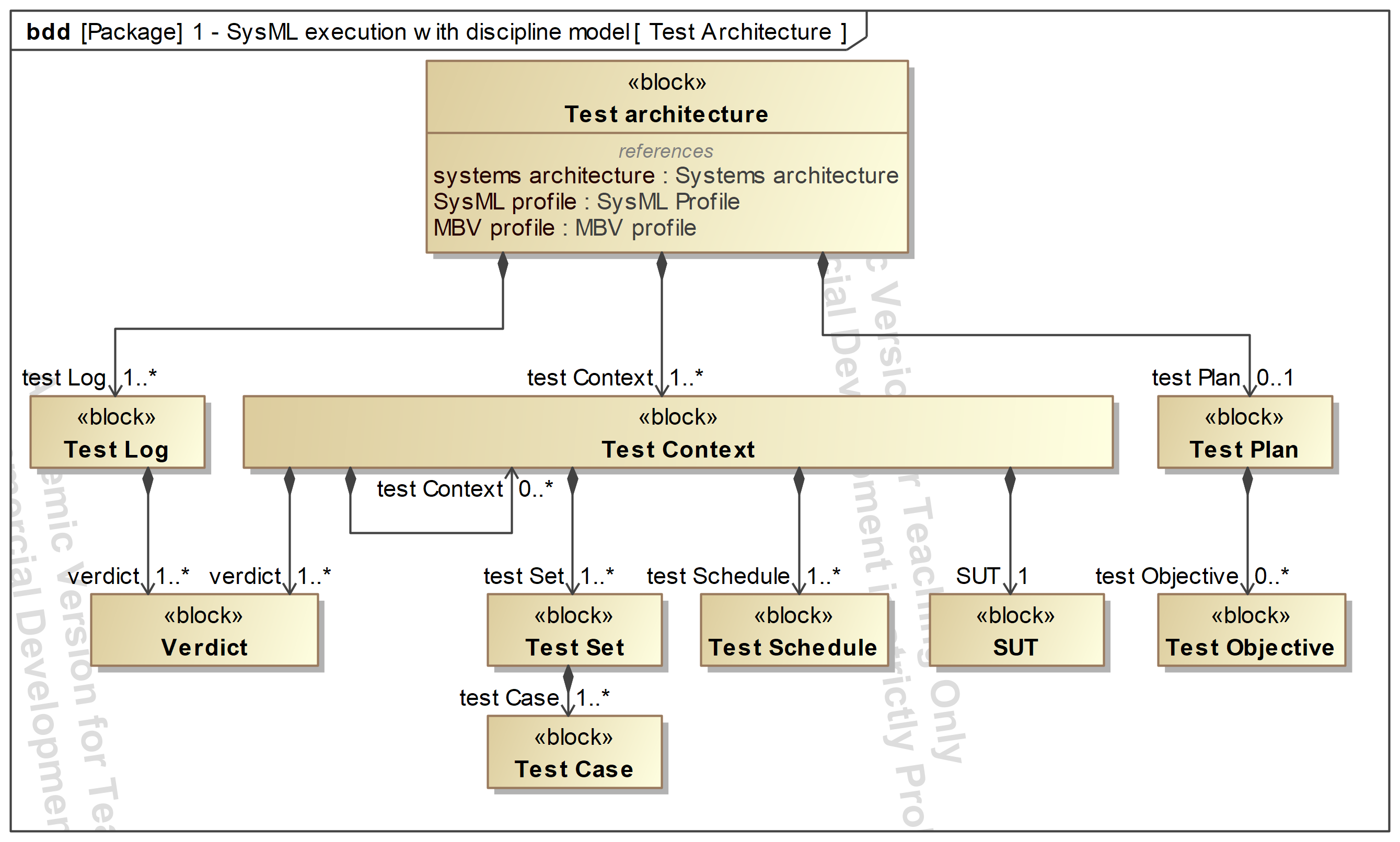}
  \caption{Logical test architecture \ac{BDD} showing \ac{MBV} profile elements and their correspondence to UTP, with departures from UTP noted for test cases, verdicts, and test logs}
  \label{fig:logical-bdd}
\end{figure}

The logical test architecture, shown in Figure~\ref{fig:logical-bdd}, closely mirrors the UTP element structure. The principal departures from UTP are that test cases and verdicts use pre-existing SysMLv1 elements rather than custom stereotypes, and that test logs are implemented as instance specifications rather than a stereotype, as prototyping showed that a stereotype was unsuitable for capturing execution results at runtime.

\subsection{Implementation}
\label{sec:implementation}

\begin{figure}[t]
  \centering
  \includegraphics[width=1\columnwidth]{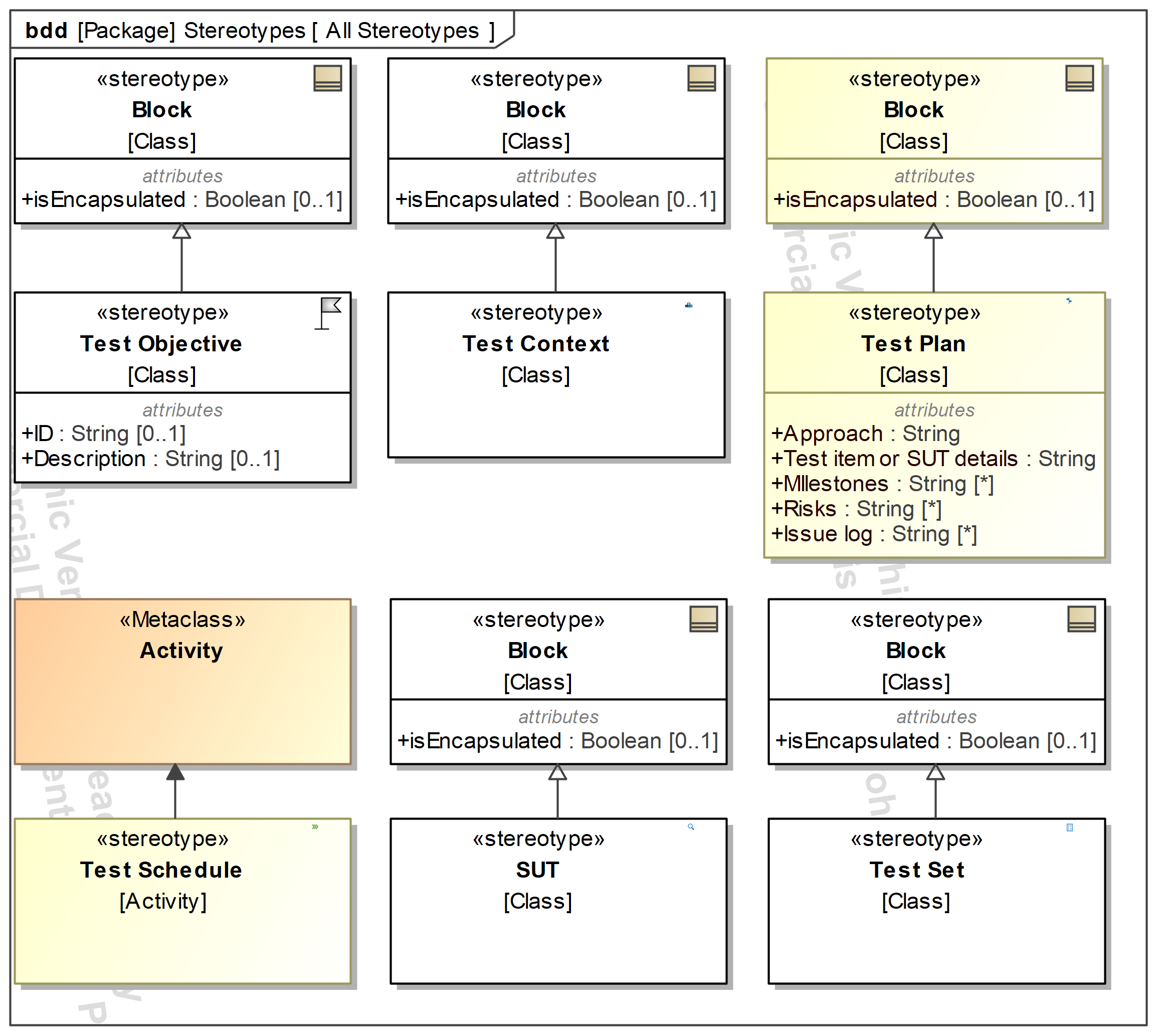}
  \caption{\ac{MBV} profile stereotype diagram as implemented in \ac{Magic SoS}, showing custom stereotype icons integrated into the tool browser and menu system}
  \label{fig:magic-sos}
\end{figure}

The \ac{MBV} profile, shown in Figure~\ref{fig:magic-sos} for Magic SoS, was implemented in both \ac{Magic SoS} and Rhapsody. The custom stereotype icons were integrated into each tool's menu options, tool panels, and model browser, enabling a systems modeller to locate and distinguish test artefacts at a glance. This toolchain-specific integration is not required for the profile to function, since stereotypes can be applied manually; it reduces friction for practitioners but does not affect portability. The profile shares significant commonality with UTP, with two deliberate differences: test logs are implemented as instance specifications to record execution results directly at runtime, and test sets absorb the test component concept, reducing the modelling burden on the practitioner.

\subsection{Demonstration}
\label{sec:demonstration}

The demonstration follows the \ac{MBV} process end-to-end to produce verification evidence and show the artefacts created at each process step. \ac{Magic SoS} is used as the primary toolchain for the full demonstration; Rhapsody examples are shown to assess the applicability of the approach across tools and to identify implementation differences.

\subsubsection{System under test}

The SUT is a Simulink model incorporating Stateflow elements, integrated with the SysML toolchains via \ac{FMU}s. The Simulink model multiplies an input signal by five at two output ports, up to a saturation limit of 25, and includes a Stateflow \ac{STM} to demonstrate message-based verification alongside the signal-based tests. The \ac{FMU} provides the co-simulation interface through which the \ac{MBV} process exercises and observes the SUT.

\subsubsection{\ac{Magic SoS}}

\begin{figure*}[t] 
  \centering
  \includegraphics[width=0.9\textwidth]{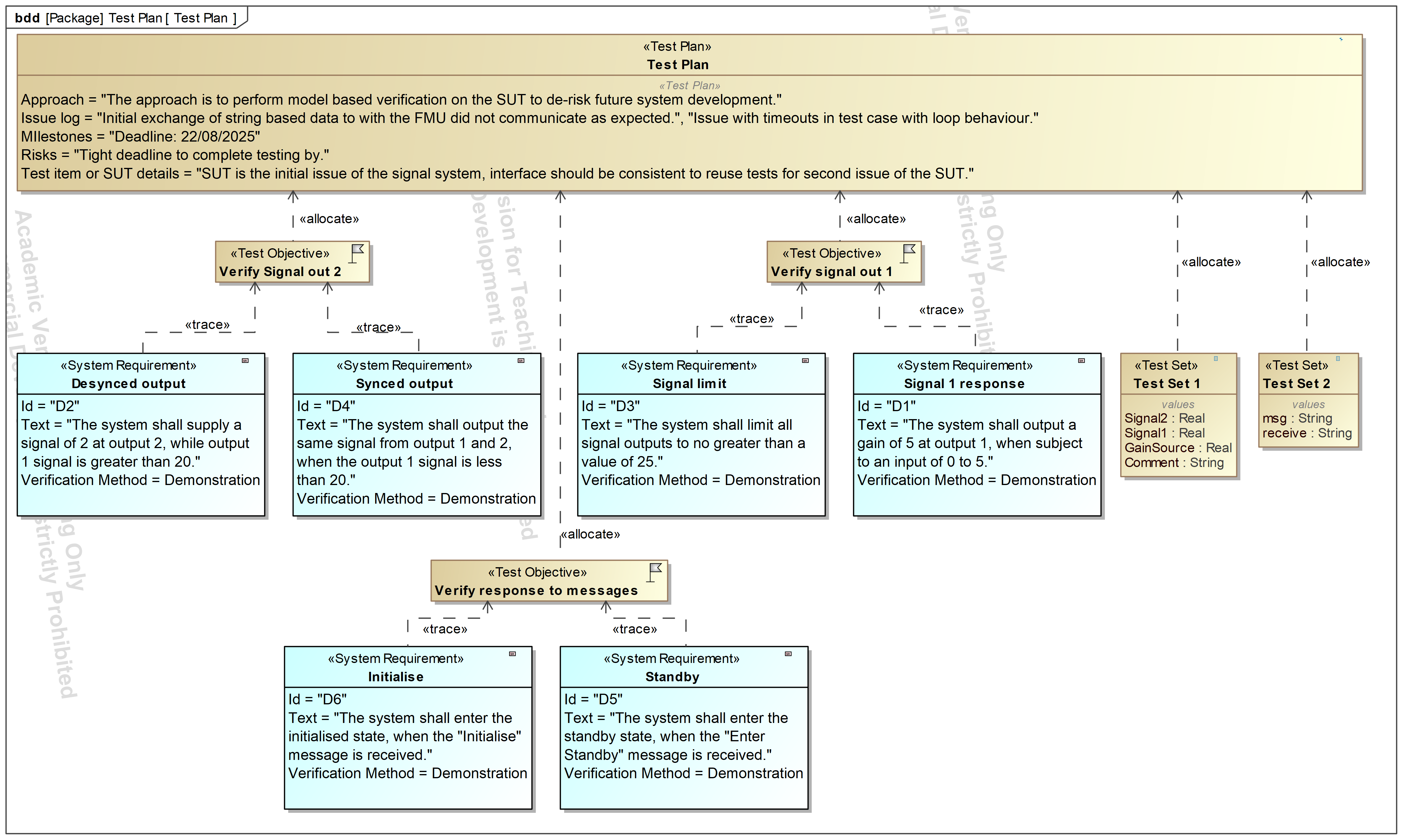} 
  \caption{Test plan artefact in \ac{Magic SoS}, showing tagged value fields for test planning information, allocated test objectives, and requirement traceability links}
  \label{fig:test-planning}
\end{figure*}

The first artefact created is the test plan, shown in Figure~\ref{fig:test-planning}. Test planning information is captured in the tagged values of the test plan stereotype; test objectives are then defined and allocated to the plan, and the requirements to be verified by \ac{MBV} are traced to those objectives, establishing traceability from requirement to test plan. Test sets created later in the process can also be allocated to the test plan.

\begin{figure}[t] 
     \centering
     \begin{subfigure}[b]{0.8\columnwidth} 
         \centering
         \includegraphics[width=0.95\linewidth]{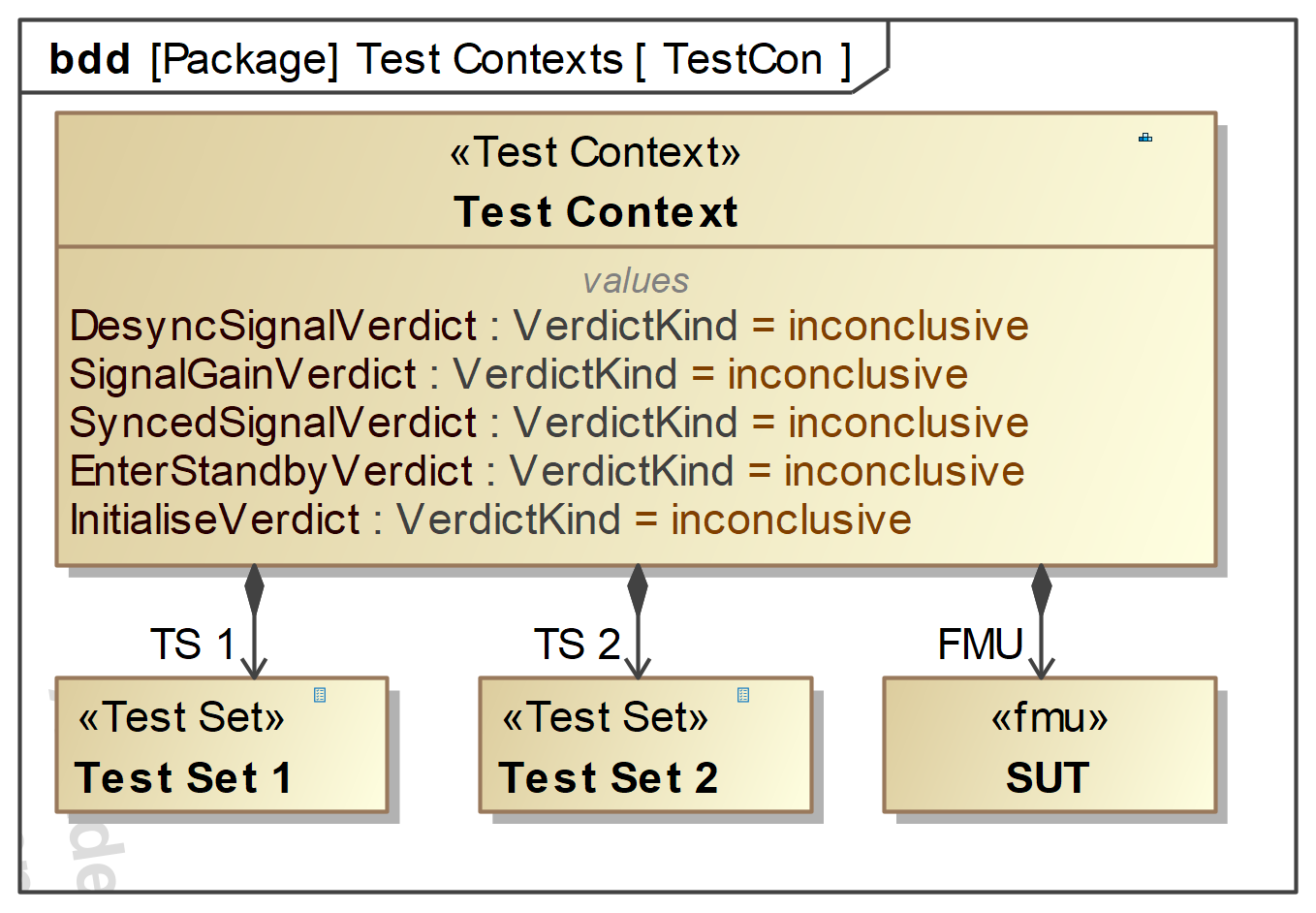} 
         \caption{Test context \ac{BDD} showing the test context block containing the test sets and the imported \ac{FMU} block representing the SUT}
         \label{fig:test_context_bdd}
     \end{subfigure}
     
     \vspace{1.5em} 
     
     \begin{subfigure}[b]{0.8\columnwidth} 
         \centering
         \includegraphics[width=0.95\linewidth]{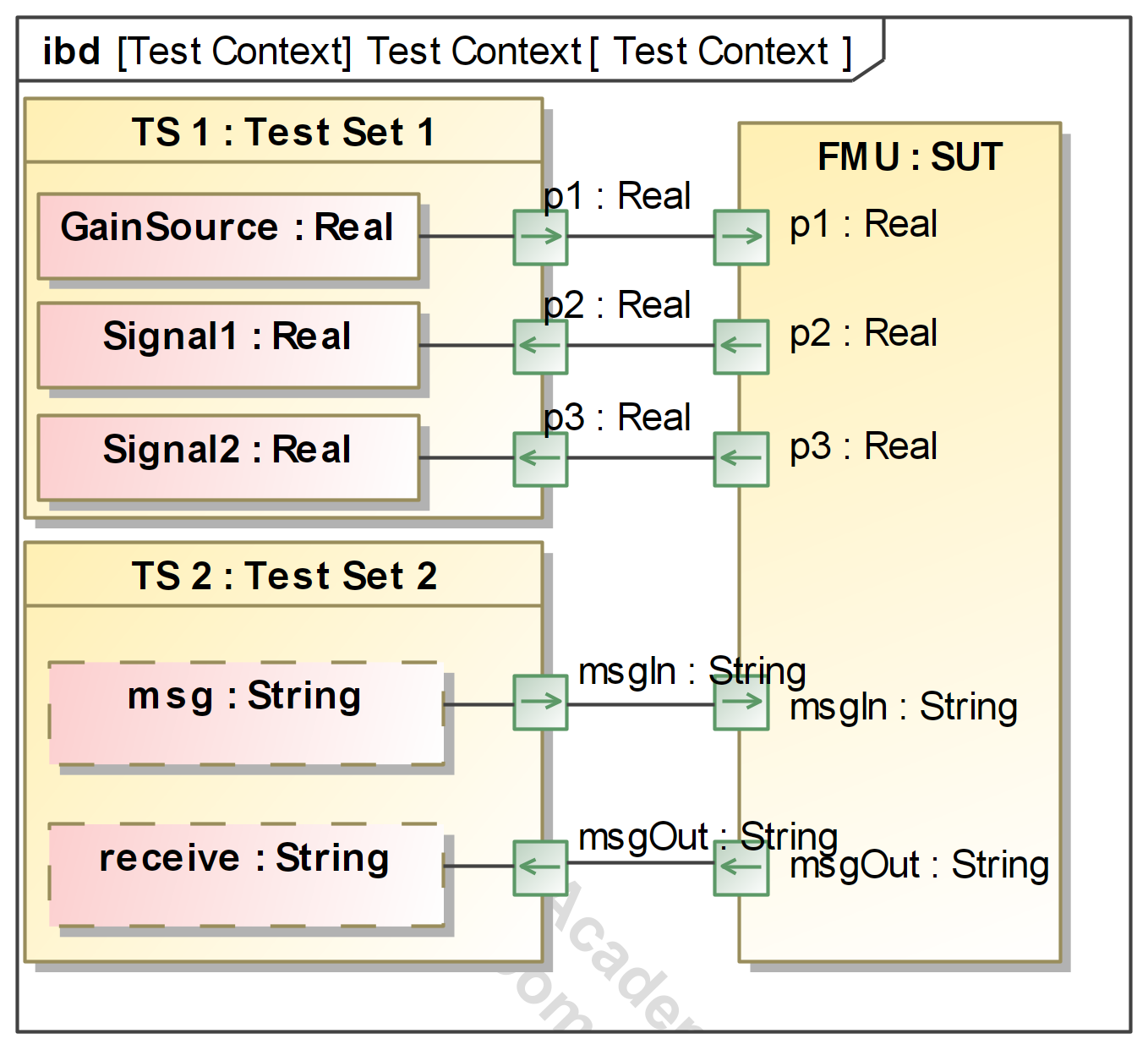}
         \caption{Test context \ac{IBD} defining the interface connections between the test set parts and the SUT \ac{FMU} ports}
         \label{fig:test_context_ibd}
     \end{subfigure}
     \caption{Test context structure: the \ac{BDD} establishes the block hierarchy and the \ac{IBD} defines the data flow interface between the test sets and the SUT}
     \label{fig:total_figure}
\end{figure}

Test cases are created by copying the template element, which prepopulates the required fields and provides default behaviour elements for verdict outcomes and a timeout that triggers an error verdict, then a verify relationship to the corresponding requirement is added. An existing \ac{AD} can also be converted directly into a test case by applying the test case stereotype to it, satisfying SN4.2. Once the Simulink model has been imported as a \ac{FMU} block, the test context structure is assembled by adding test sets and the SUT to a test context block, with the interface between the parts defined on an \ac{IBD}. The resulting \ac{BDD} and \ac{IBD} are shown in Figure~\ref{fig:total_figure}.

\begin{figure}[ht!]
  \centering
  \includegraphics[width=1\columnwidth]{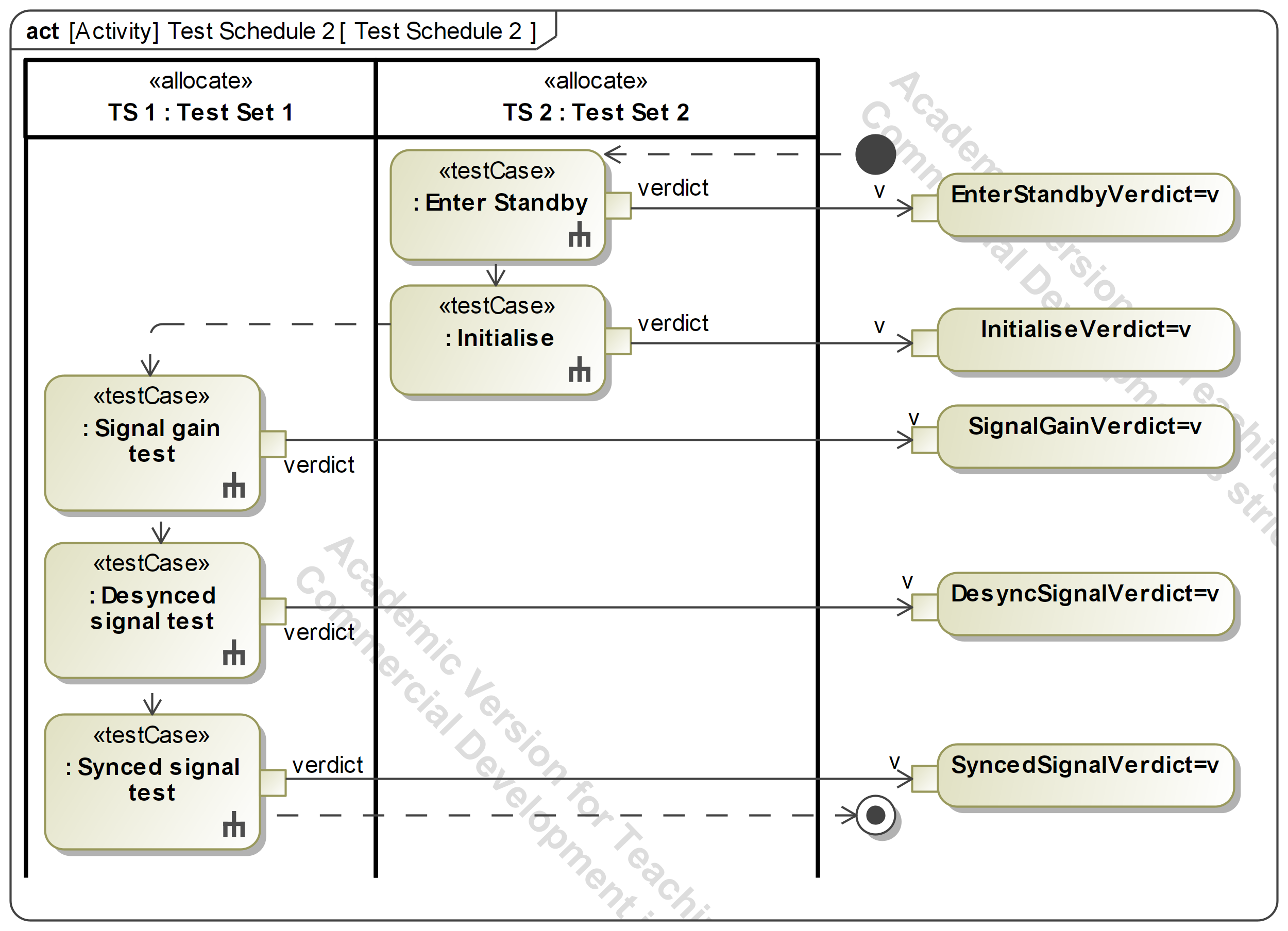}
  \caption{Test schedule \ac{AD} using call behaviour actions and control flows to sequence test cases and collect verdicts into the test context}
  \label{fig:test-schedule}
\end{figure}

Test cases are defined as activities within the test sets using executable \ac{AD}s, selected because they are more broadly supported for execution across SysMLv1 tools than \ac{SD}s. The test schedule is then defined as a separate \ac{AD} using call behaviour actions and control flows to sequence the test cases, as shown in Figure~\ref{fig:test-schedule}. As each test case completes, the schedule retrieves the result and stores the verdict as a value property within the test context. Where test cases naturally follow one another, the schedule can exploit prior test inputs to reduce modelling effort; a dependency matrix template included in the profile library makes inter-case dependencies explicit.

\begin{figure}[ht!]
  \centering
  \includegraphics[width=1\columnwidth]{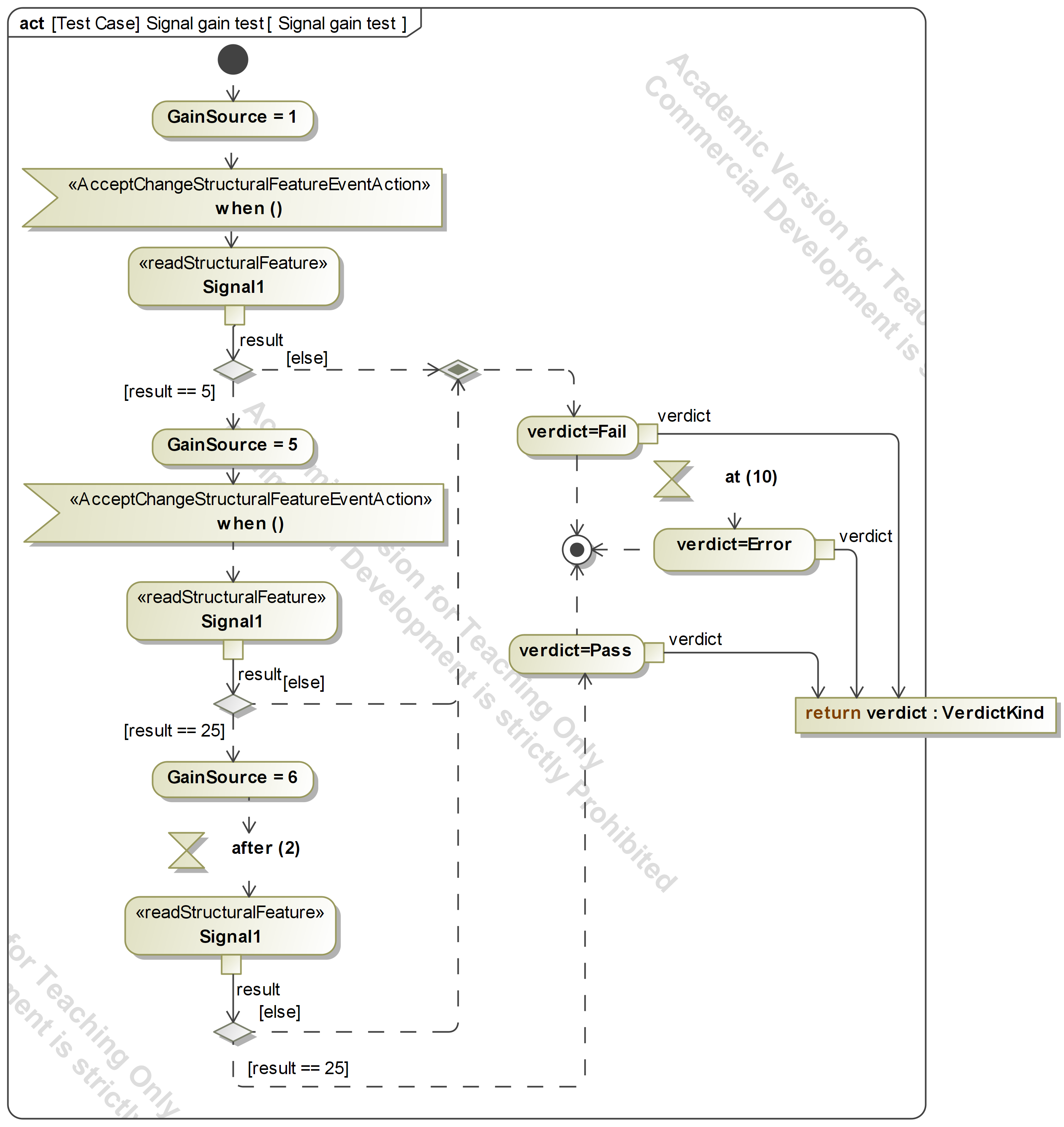}
  \caption{Signal gain test case \ac{AD} verifying two requirements: a ubiquitous output limit requirement assessed via a time event, and an event-driven gain requirement assessed via an accept change event with decision node guards}
  \label{fig:signal-gain}
\end{figure}

In applying the \ac{EARS}-to-SysML mapping from Section~\ref{sec:system-requirements-test-cases} and Table~\ref{tab:ears-mapping} to test case definition, one departure from the mapping was necessary: WHEN clause triggers, which the mapping associates with send actions, were not required in practice because the flow ports transmit automatically whenever a linked value property changes. Accept change events were therefore used to make the test case wait for a system response before evaluating it, with a decision node and guards performing the assessment. For the signal gain test shown in Figure~\ref{fig:signal-gain}, a time event rather than an accept change event was used to assess the output limit requirement, since that output reaches its maximum without a further value change to detect; an additional time event provides a timeout to flag a failure if no result is returned within the expected window. The two requirements verified by this test case and their \ac{EARS} syntax annotations are shown in red on Figure~\ref{fig:signal-gain}.

\begin{figure}[ht!]
  \centering
  \includegraphics[width=\columnwidth]{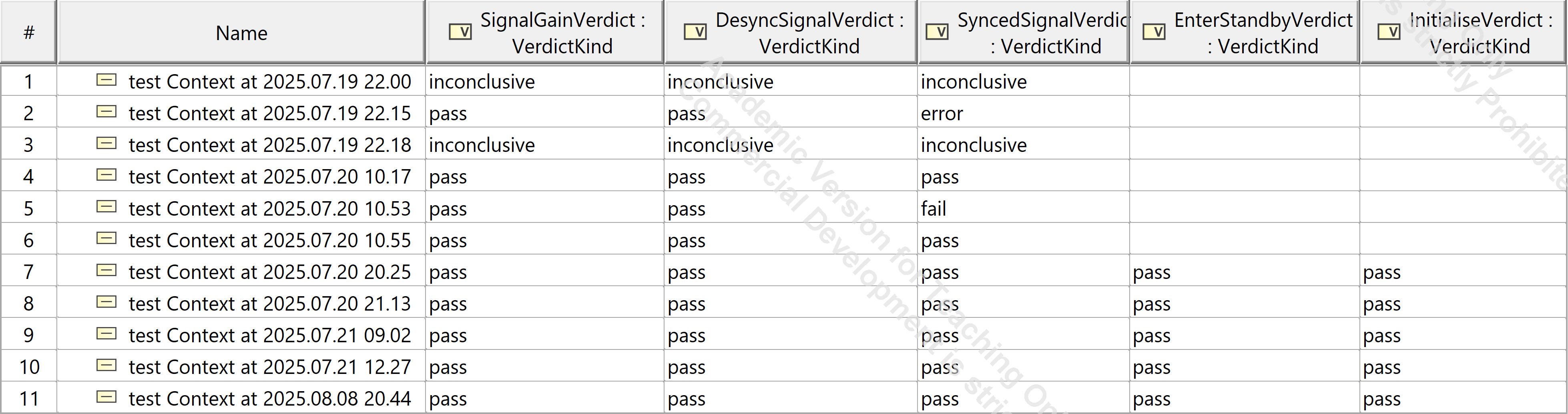}
  \caption{Test log instance specifications recording the verdict and parameter values for each completed test case}
  \label{fig:test-logs}
\end{figure}

Where a test case depends on a preceding one having already set inputs, a read structural feature block can assess the previously received value without re-sending it. This pattern is useful for SUTs that require an initialisation or startup sequence, which can alternatively be called into each dependent test case with a call behaviour action. After all test cases have executed, results are stored as instance specifications, as shown in Figure~\ref{fig:test-logs}. Table and matrix views, which are not part of the SysML standard but are widely available across SysMLv1 tools, can be used to support the process with visibility of coverage and dependencies.

\subsubsection{Rhapsody}

Rhapsody supported the same overall \ac{MBV} workflow but with three notable differences. First, model execution behaved slightly differently between the two tools, which was expected given that SysMLv1 does not define an execution standard. Second, Rhapsody test case elements would not compile for execution if they contained accept change events or timeout events; as a workaround, test cases were implemented as callable activities and timeouts were managed externally as state transitions in a \ac{STM} test schedule, an approach that more closely mirrors the \ac{Magic SoS} activity-based implementation. Third, execution results could not be saved directly to instance specifications during simulation; achieving this natively requires API access to retrieve values from the execution instance during the simulation run, though exporting results to a file is a simpler alternative. Structural elements, requirement traceability, and table and matrix views were consistent with \ac{Magic SoS}.

\subsection{Verification and validation}
\label{sec:verification-and-validation}

Table~\ref{tab:consolidated} records the verification and validation outcome for each stakeholder need against both toolchains. Of the nineteen needs assessed, eighteen pass for \ac{Magic SoS} and seventeen for Rhapsody. SN4.1, the ability to generate a test case directly from a requirement, failed for both tools as this is not achievable without tool-specific automation; the template-based approach provides an equivalent outcome with manual initiation. SN6.1, the recording of verification results, passed for \ac{Magic SoS} through instance specification population during simulation but failed for Rhapsody as a tool specific automation would be required.

\section{Discussion}
\label{sec:discussion}

The synthesised \ac{MBV} process defined in Section~\ref{sec:literature-review}, drawing on ISO 15288, ISO 24641, and the INCOSE \ac{SEH}, provided a suitable framework for the demonstration reported in Section~\ref{sec:systems-engineering-dev}. The process addressed each of the validated stakeholder needs and supported the end-to-end workflow from test planning through to the return of results to the SysML model. The three-phase structure of prepare, perform, and manage results proved sufficient to organise the toolchain demonstration, and the \ac{AD}s used to specify the process translated naturally into the modelling artefacts created in both toolchains.

The most significant contribution of the work is the extension of \ac{MBV} scope beyond the parametric focus of the existing literature. Prior approaches have verified performance properties by driving parametric diagrams with discipline-specific model outputs, a technique that can assess whether a system achieves a numerical target but cannot verify temporal ordering, timeout behaviour, or state-dependent responses. By using SysML \ac{AD} constructs, specifically decision nodes, time events, and accept change events, the demonstrated approach can verify these behavioural and interface properties directly from test cases defined in the SysML model. This extends the class of requirements that \ac{MBV} can address without requiring any addition to the verification infrastructure. The approach to test case definition was also distinctive in comparison to existing \ac{AD} based examples in the literature, which rarely made use of decision logic or timeout handling. Systems engineering practice can benefit from incorporating these constructs, borrowed from the UTP tradition of software testing, into \ac{MBV} workflows that target more complex discipline-specific models. The present work automates the execution of a given verification strategy rather than its selection; the complementary problem of determining which verification activities to perform and in what sequence has been formalised separately as a set-based design and decision problem \citep{Salado2018Mathematical,Salado2019Elemental}, and lies outside the scope of this paper.

The \ac{MBV} profile supported the process by providing domain-specific language that guides the system modeller through the workflow and by supplying template elements that accelerate test architecture construction. Compared to UTP, the profile is deliberately simpler: no additional stereotypes were needed beyond those required for automation, because the SysML tools provided equivalent elements natively. Where existing profile implementations in the literature are more elaborate, this is typically because their API or model transformation-based solutions require additional metadata to drive the integration. This project avoided API-specific automation in favour of library templates precisely to preserve tool agnosticism, a design decision that makes adoption more straightforward in collaborative industrial environments where teams do not share a common toolchain. One departure from UTP conventions was the decision to merge the test component concept with test sets. This reduces the number of model elements a practitioner must create and maintain, at no cost to the verification information captured, and aligns better with the way test artefacts are organised in current SysML tool browsers.

The demonstration confirmed that the logical architecture of the process and profile is applicable across the majority of SysMLv1 tools, but also exposed a specific limitation: a completely tool-agnostic implementation is not achievable with SysMLv1, because tool vendors have implemented both the SysML specification and model execution in inconsistent ways. In the demonstration, this manifested as differences in how test case elements behaved in \ac{Magic SoS} and Rhapsody, requiring manual adjustment after model transfer rather than seamless portability. In particular, Rhapsody was unable to save execution results directly to instance specifications during simulation, meaning that SN6.1 passed for \ac{Magic SoS} but failed for Rhapsody. These findings are consistent with the known fragmentation of SysMLv1 tool implementations and reinforce the case for either API-based result capture or file export as fallback mechanisms, both of which are feasible within the current architecture.

Two further technical limitations of the implementation merit attention for industrial adoption. First, the \ac{FMI} v1 and v2 standards used for model exchange support flat data types but not structured data types, for Rhapsody and Magic SoS, respectively. Real engineering projects frequently represent interface signals as structured types, so this constrains the approach when message-based verification is required. Wrapper functions that flatten and reconstruct structured data at the exchange boundary are a viable mitigation, though they introduce processing overhead. \ac{FMI} v3 may alleviate this constraint if it becomes available in the relevant toolchains. Second, a discipline-specific model exported as an \ac{FMU} does not produce numerically identical results to the original simulation \citep{Cederbladh2024Earlya}. For verification purposes, this requires careful consideration, since test case verdicts depend on the fidelity of the discipline-specific model output. The magnitude of the discrepancy will vary with model complexity and solver configuration, and establishing acceptable tolerance bounds should be part of any industrial deployment of this approach.

The SysMLv2 situation warrants a brief remark. SysMLv2 is expected to address the inconsistent tool implementation problem through a standardised API and a standardised textual notation, which in principle could make the portability of test cases between tools more reliable than is possible with SysMLv1. However, SysMLv2 does not include a standard for model execution, so the execution inconsistencies observed in the SysMLv1 demonstration are likely to recur unless a separate execution standard is defined. If the SysMLv2 API is implemented consistently across tools, an alternative architecture in which a third-party testing tool reads and writes the SysML model through the API may prove more suitable than native model execution. These considerations are left as directions for future work.

\section{Conclusions}
\label{sec:conclusions}

This paper has presented a tool-agnostic \ac{MBV} process and supporting SysML profile for the automated verification of discipline-specific models from SysML test cases, with results returned to the SysML model and traceable to requirements. The process was synthesised from ISO\slash IEC\slash IEEE~15288, ISO/IEC/IEEE~24641, and the INCOSE \ac{SEH}, and the profile was designed around capabilities common to current SysMLv1 tools to avoid dependence on any single vendor's implementation. End-to-end demonstration in Magic System of Systems Architect and IBM Engineering Systems Design Rhapsody confirmed that the logical architecture is broadly applicable across SysMLv1 toolchains, though the demonstration also revealed that a fully seamless implementation is not achievable under SysMLv1 due to inconsistent tool interpretations of the specification. The most significant contribution is the extension of \ac{MBV} scope to include behavioural and interface requirements alongside performance requirements, demonstrated through SysML \ac{AD} constructs that existing parametric-only approaches cannot replicate. Together, these results show that meaningful progress on both the portability and the scope limitations of current SysML-based \ac{MBV} practice is achievable within the current generation of SysMLv1 toolchains.

Two directions for further work follow from these findings. The first concerns the maturation of the SysMLv1 implementation. The verification scope demonstrated here is necessarily bounded by the test cases constructed; a broader set of examples is needed to establish full coverage of requirement types and to expose any constructs for which the \ac{AD} approach proves insufficient. Resolving the structured data type constraint imposed by \ac{FMI} v2 is a prerequisite for testing realistic message\-/based interfaces, and should be addressed either via \ac{FMI} v3 adoption or through data flattening strategies. Establishing this more complete SysMLv1 baseline also matters because it would provide the clearest possible point of comparison for the second direction. The second concerns the transition to SysMLv2. SysMLv2 addresses the portability problem differently, through a standardised \ac{API} rather than through model execution consistency, but it does not define an execution standard, so the behavioural verification capability demonstrated here would require a different architectural approach. An \ac{API}-driven architecture in which a third-party testing tool reads and writes the SysML model directly is the most promising candidate, but this remains to be designed and evaluated. These two directions together define a research path that leads from the present work toward an industrially viable \ac{MBV} capability.

\section*{Acknowledgement}
Daniel Marley would like to thank Gareth Neely, Steve McQueen, and Ranjit Ravindranath for acting as stakeholders and providing valuable feedback. The authors also thank Steve Hinsley and Michael Henshaw for their valuable comments.

\bibliographystyle{plainnat}
\bibliography{references}

\section*{Appendix 1 -- Acronyms and Terms}

\nopagebreak 
\begin{acronym}[Magic SoS]
    \setlength{\itemsep}{0.0ex}  
    \setlength{\parsep}{0pt}     
    
    \acro{AD}{Activity Diagram}
    \acro{API}{Application Programming Interface}
    \acro{BDD}{Block Definition Diagram}
    \acro{DSR}{Design Science Research}
    \acro{EARS}{Easy Approach to Requirements Syntax}
    \acro{FMI}{Functional Mock-up Interface}
    \acro{FMU}{Functional Mock-up Unit}
    \acro{IBD}{Internal Block Diagram}
    \acro{Magic SoS}{Magic Systems of Systems Architect}
    \acro{MBSE}{Model-Based Systems Engineering}
    \acro{MBV}{Model-Based Verification}
    \acro{Par}{Parametric diagram}
    \acro{SD}{Sequence Diagram}
    \acro{SEH}{Systems Eng. Handbook}
    \acro{STM}{State machine}
    \acro{SysML}{Systems Modeling Language}
    \acro{UC}{Use Case}
    \acro{UML}{Unified Modeling Language}
    \acro{UTP}{UML Testing Profile}
\end{acronym}

\section*{Appendix 2 -- Stakeholder needs and V\&V results}
\label{app:vv}
\addcontentsline{toc}{section}{Appendix 2 -- Stakeholder needs and verification results}

\begin{widelongtable}
\small 
\centering
\setlength{\tabcolsep}{2.5pt} 
\renewcommand{\arraystretch}{1.0} 
\footnotesize 

\begin{xltabular}{\textwidth}{|p{0.75cm}|>{\hsize=1.15\hsize}X|>{\hsize=0.85\hsize}X|}
\caption{Consolidated stakeholder needs, MoSCoW prioritisation, and verification results}\label{tab:consolidated} \\
\hline
\rule{0pt}{3.5ex}\rule[-1.5ex]{0pt}{0pt}\small\textbf{Need} & 
\small\textbf{Specification \hfill [MoSCoW]} & 
\small\textbf{V\&V Statement \hfill [Result]} \\
\hline
\endfirsthead
\multicolumn{3}{l}{\itshape Table~\ref{tab:consolidated} continued} \\
\hline
\rule{0pt}{3.5ex}\rule[-1.5ex]{0pt}{0pt}\small\textbf{Need} & 
\small\textbf{Specification \hfill [MoSCoW]} & 
\small\textbf{V\&V Statement \hfill [Result]} \\
\hline
\endhead
\hline
\multicolumn{3}{r}{\itshape continued on next page} \\
\endfoot
\hline
\endlastfoot
\multicolumn{3}{|l|}{\rule{0pt}{3.5ex}\rule[-1.5ex]{0pt}{0pt}\small\textbf{SN1 Model customisation}} \\
\hline
SN1.1 & The user needs to extend the SysML model functionality to automate testing of discipline-specific models. \hfill \textbf{[Could]} & \ac{MBV} profile provides additional required types. \hfill \textbf{[Pass]} \\
\hline
SN1.2 & The user needs to extend the SysML tool functionality to automate testing of discipline-specific models. \hfill \textbf{[Could]} & No tool extension required. \hfill \textbf{[Pass]} \\
\hline
\multicolumn{3}{|l|}{\rule{0pt}{3.5ex}\rule[-1.5ex]{0pt}{0pt}\small\textbf{SN2 Test management}} \\
\hline
SN2.1 & The user needs to create a package containing all elements needed for the SUT. \hfill \textbf{[Must]} & Demonstrated by Test Context structure. \hfill \textbf{[Pass]} \\
\hline
SN2.2 & The user needs to link to or manage test plans within the model. \hfill \textbf{[Should]} & \ac{MBV} profile meets need with additional types, where an external test plan is undefined. \hfill \textbf{[Pass]} \\
\hline
SN2.3 & The user needs to define the order in which tests will run. \hfill \textbf{[Must]} & Demonstrated via test schedule. \hfill \textbf{[Pass]} \\
\hline
SN2.4 & The user needs to create a test set. \hfill \textbf{[Must]} & Demonstrated via Test Set element. \hfill \textbf{[Pass]} \\
\hline
SN2.5 & The user needs to link test cases to requirements. \hfill \textbf{[Must]} & Demonstrated via SysML traceability links. \hfill \textbf{[Pass]} \\
\hline
SN2.6 & The user needs to generate requirement coverage reports on coverage of requirements by test cases. \hfill \textbf{[Must]} & Covered by SysML relationships and table or matrix views. \hfill \textbf{[Pass]} \\
\hline
\multicolumn{3}{|l|}{\rule{0pt}{3.5ex}\rule[-1.5ex]{0pt}{0pt}\small\textbf{SN3 Interface management}} \\
\hline
SN3 & The user needs to define the interface between the systems and discipline model. \hfill \textbf{[Must]} & Defined by \ac{IBD} and \ac{FMI}. \hfill \textbf{[Pass]} \\
\hline
\multicolumn{3}{|l|}{\rule{0pt}{3.5ex}\rule[-1.5ex]{0pt}{0pt}\small\textbf{SN4 Test case definition}} \\
\hline
SN4.1 & The user needs to create a test case from a requirement. \hfill \textbf{[Should]} & Not possible directly. \hfill \textbf{[Fail]} \\
\hline
SN4.2 & The user needs to create a test case from a behavioural diagram. \hfill \textbf{[Should]} & Demonstrated with activity diagrams. \hfill \textbf{[Pass]} \\
\hline
\multicolumn{3}{|l|}{\rule{0pt}{3.5ex}\rule[-1.5ex]{0pt}{0pt}\small\textbf{SN5 Verification execution}} \\
\hline
SN5.1 & The user needs to execute a test case. \hfill \textbf{[Must]} & Via test schedule. \hfill \textbf{[Pass]} \\
\hline
SN5.2 & The user needs to execute a test package. \hfill \textbf{[Should]} & Via test schedule. \hfill \textbf{[Pass]} \\
\hline
SN5.3 & The user needs to execute a test set. \hfill \textbf{[Should]} & Via test schedule. \hfill \textbf{[Pass]} \\
\hline
\multicolumn{3}{|l|}{\rule{0pt}{3.5ex}\rule[-1.5ex]{0pt}{0pt}\small\textbf{SN6 Record verification outputs}} \\
\hline
SN6.1 & The user needs to record outcomes of verification actions. \hfill \textbf{[Must]} & Results recorded in instance specifications in \ac{Magic SoS}; Rhapsody cannot do this directly. \hfill \textbf{[Pass / Fail]} \\
\hline
SN6.2 & The user needs to generate witness scenarios to visualise test results. \phantom{filler} \hfill \textbf{[Should]} & Via \ac{SD} records. \hfill \textbf{[Pass]} \\
\hline
SN6.3 & The user needs to generate model coverage reports. \hfill \textbf{[Could]} & Covered by SysML relationships and table or matrix views. \hfill \textbf{[Pass]} \\
\hline
SN6.4 & The user needs to generate requirement coverage reports on coverage by test results. \hfill \textbf{[Should]} & Covered by SysML relationships and table or matrix views. \hfill \textbf{[Pass]} \\
\hline
SN6.5 & The user needs to export results to other formats. \hfill \textbf{[Could]} & Via tool report generation capabilities. \hfill \textbf{[Pass]} \\
\end{xltabular}
\end{widelongtable}

\end{document}